\documentclass[aps,prd,superscriptaddress,nofootinbib,preprint]{revtex4}

\usepackage{float}
\usepackage{amsmath,amssymb}
\usepackage[dvips]{graphicx}
\DeclareGraphicsExtensions{.jpg,.mps,.pdf,.png,.eps} 
\usepackage{color}

\newcommand{\non}{\nonumber}
\newcommand{\be}{\begin{equation}}
\newcommand{\ee}{\end{equation}}                  
\newcommand{\bea}{\begin{eqnarray}}
\newcommand{\eea}{\end{eqnarray}}

\begin{document}

 
\title{Mimicking dark matter in Horndeski gravity.}


\author{Massimiliano Rinaldi}
\email{massimiliano.rinaldi@unitn.it}
\affiliation{Dipartimento di Fisica, Universit\`a di Trento,\\  Via Sommarive 14, 38123 Povo (TN), Italy}
\affiliation{TIFPA - INFN,\\  Via Sommarive 14, 38123 Povo (TN), Italy}

\begin{abstract}
Since the rediscovery of Horndeski gravity, a lot of work has been devoted to the exploration of its properties, especially in the context of dark energy. However, one sector of this theory, namely the one containing the coupling of the Einstein tensor to the kinetic term of the scalar field, shows some surprising features in the construction of black holes and neutron stars. Motivated by these new results, I explore the possibility that this sector of Horndeski gravity can mimic cold dark matter at cosmological level and also explain the flattening of galactic rotation curves. I will show that, in principle, it is possible to achieve both goals with at least two scalar fields and a minimal set of assumptions.
\end{abstract}

\maketitle

\section{Introduction}

\noindent In the last decade a lot of interest has been devoted to extended theories of gravity. Generally speaking, these take the form of a generalisation of the Einstein-Hilbert action (such as $f(R)$ \cite{reviewfR}) or of a generalisation of tensor-scalar theories (such as Horndeski gravity \cite{Horndeski}). The common denominator of all these models is the avoidance of the Ostrogradski instability (at least at the background level) by carefully engineering the action so that the equations of motion are differential equations of second order. 

There are several motivations that push research beyond standard general relativity (GR). On one hand the need for a formally consistent quantisation seems to require the extension of GR with higher order terms. On the other hand, phenomena like dark energy and dark matter can be explained by modified GR (see e.g. \cite{odint}) or by taking in account the dynamics of the Higgs sector (see eg \cite{YMDE} for dark energy). Repeated experimental failures to explain dark matter with some unknown fundamental particle keep an open door to the possibility that dark matter is a modification of gravity rather than a hidden sector of the Standard Model, according to the old ``modified Newton dynamics'' (MOND) spirit \cite{mondreview}. Generally, the entire $\Lambda CDM$ model is under the pressure of a number of investigations and data of increasing accuracy  \cite{review}. There are great expectations from experiments such as Euclid that should shed some light on the dark side of the Universe \cite{euclid}.

Of course, whenever GR is modified at some scale one should check the consistency of the new theory at all scales, in particular at those where experimental tests are available. Currently, the most accurate data on GR come from Solar System measurements, in particular the Cassini doppler experiment \cite{cassini}, but other tests at larger scale are equally important.

Any modified theory of gravity that does not violate the very stringent Solar System constraints should  be taken seriously and tested at other scales. For example, the gravitational model of Higgs inflation \cite{Bezrukov:2010jz} was studied in the context on compact object solutions \cite{HiggsNS} showing that it is negligibly different from GR around stellar compact objects. 

These tests are particularly relevant in the case of a modified theory of gravity that  matches \emph{exactly} GR outside a compact object thus passing by default all Solar System tests. As strange as it may sound such a theory exists and has been studied in the context of static and slowly rotating neutron stars in \cite{HornNS} (for further extensions see \cite{HornNSext}). The model is based on a subsector of Horndesky gravity  and it is characterised by the coupling of the Einstein tensor to the kinetic term of the scalar field. In the absence of a scalar field potential, this model is invariant under shifts of the scalar field. If also the usual kinetic term for the scalar field is negligible, then the metric outside a compact object matches the Schwarzschild one even though the scalar field does not vanish anywhere.

The coupling between the Einstein tensor and the kinetic term of the scalar field arises in Horndeski gravity \cite{Horndeski} and has interesting properties besides keeping the equations of motion of second of order. When quantum effects are taken in account in GR it is common to add higher order terms to the Einstein-Hilbert Lagrangian, constructed out of contractions of the Riemann tensor. In general all these terms are severely suppressed by the Planck mass and become relevant only in the inflationary Universe. However, when one considers the kinetic term of the scalar field of the form $(\alpha g_{\mu\nu}-\eta G_{\mu\nu})\partial^{\mu}\psi\partial^{\nu}\psi$, where $G_{\mu\nu}$ is the Einstein tensor, one opens up the possibility to add high-order corrections (in the form of second derivatives of the metric) to the Einstein-Hilbert Lagrangian of a tensor-scalar theory that are not suppressed by the Planck mass and can modify physics at lower energies. When the curvature is very small, the second term is negligible anyway thus it cannot affect, say, the Standard Model physics. This implies the possibility that this kind of coupling exists for all fields of the Standard Model. In turn, this would implies a coupling of the Einstein tensor with the matter part of the Lagrangian. I  will leave for future work this intriguing possibility. Here, I  just consider the coupling to a single scalar field and investigate the modifications to the dynamics at both galactic and cosmological scales.

The plan of the paper is the following. In the next section I  lay down the main equations of the system. In secs.\ \ref{sec3} and \ref{sec4}  I  study two scenarios, one in where the scalar field acts as dark matter on cosmic scales and another where it appears as an extra matter  component. In sec.\ \ref{sec5} I  show how this model can justify the flattening of galactic rotation curves. I  draw some conclusions in sec.\ \ref{sec6}. In the following I  will use the mostly plus signature.

\section{Cosmological dynamics}\label{sec2}

\noindent The models of neutron stars studied in  \cite{HornNS} are based on  an action inspired by Horndeski gravity and given by
\begin{equation}
S=\int d^{4}x\sqrt{-g}\left[\kappa(R-2\Lambda)-\frac{1}{2}\left(\alpha g_{\mu\nu}-\eta G_{\mu\nu} \right)\nabla^{\mu}\psi\nabla^{\nu}\psi \right] +S_{m}[g_{\mu\nu}]\,.\label{model}
\end{equation}
Here $\kappa=(16\pi G)^{-1}$, $\alpha$ and $\eta$ are two parameters controlling the strength of the  kinetic couplings. $S_{m}$ is the action of radiation and matter fields, which I  consider to be minimally coupled to the metric $g_{\mu\nu}$. Note that $\alpha$ is dimensionless while $\eta$ has dimensions (mass)$^{-2}$. One important feature of this model is that the shift symmetry $\psi\rightarrow\psi +\psi_0$ implies that the equation of motion for the scalar field can be written as the current conservation law.  Thus, the equations of motion are
\begin{eqnarray}
G_{\mu\nu}+\Lambda g_{\mu\nu}+H_{\mu\nu}=T_{\mu\nu}^{\rm fluid}\,, \label{eqmetric}\\
\nabla_{\mu} J^\mu  =0\,, \label{eqphi}
\end{eqnarray}
where
\begin{eqnarray}
H_{\mu\nu}&=&-\frac{\alpha}{2\kappa}\Bigg[\nabla_{\mu}\psi\nabla_{\nu}\psi-\frac{1}{2}g_{\mu\nu}\nabla_{\lambda}\psi\nabla^{\lambda}\psi\Bigg]\nonumber-\frac{\eta}{2\kappa}\Bigg[\frac{1}{2}\nabla_{\mu}\psi\nabla_{\nu}\psi R-2\nabla_{\lambda}\psi\nabla_{(\mu}\psi R_{\nu)}^{\lambda}\nonumber\\
&-&\nabla^{\lambda}\psi\nabla^{\rho}\psi R_{\mu\lambda\nu\rho}-(\nabla_{\mu}\nabla^{\lambda}\psi)(\nabla_{\nu}\nabla_{\lambda}\psi)+\frac{1}{2}g_{\mu\nu}(\nabla^{\lambda}\nabla^{\rho}\psi)(\nabla_{\lambda}\nabla_{\rho}\psi)-\frac{1}{2}g_{\mu\nu}(\square\psi)^{2}\nonumber\\
&&+(\nabla_{\mu}\nabla_{\nu}\psi)\square\psi+\frac{1}{2}G_{\mu\nu}(\nabla\psi)^{2}+g_{\mu\nu}\nabla_{\lambda}\psi\nabla_{\rho}\psi R^{\lambda\rho}\Bigg] \,,\\
J^\mu &=&  \left(  \alpha g^{\mu\nu}-\eta G^{\mu\nu}\right)\nabla_{\nu}\psi\,,\label{current}
\end{eqnarray}
and $T_{\mu\nu}^{\rm fluid}$ is the energy momentum tensor of one or more perfect fluids. When $\alpha\neq 0$, it can be reabsorbed into $\eta$ and by a redefinition of $\psi$, so it is sufficient to consider only the cases $\alpha=0,\pm 1$.  In the ``Fab Four'' language of ref. \cite{fabfour}, this Lagrangian corresponds to the ``John''  term, see also \cite{namur}.

We choose a flat Robertson-Walker metric of the form
\bea
ds^{2}=-dt^{2}+a(t)^{2}\delta_{ij}dx^{i}dx^{j}\,,
\eea
 and I  assume that the scalar field depends on the cosmic time only.
Then, since  the equations of motion will depend on $\dot\psi$ and $\ddot \psi$ only, I  define the new field  
\bea
\phi\equiv\dot\psi\,.
\eea
Note that this field has dimension (mass)$^{2}$. The Friedmann equations read
\bea
H^{2}&=&  {4\Lambda \kappa+\alpha\phi^{2}+2\rho_{r}+2\rho_{m}\over 3(4\kappa-3\eta\phi^2)}\,\label{fr1},\\\non\\\label{fr2}
\dot H&=& -{3\rho_{m}+4\rho_{r}\over 3(4\kappa-3\eta\phi^{2})}+{\phi\dot\phi(\alpha+9\eta H^{2})\over 3H(4\kappa-3\eta\phi^{2})}  \,,
\eea
where $H=\dot a/a$, $\rho_{r}$ and $\rho_{m}$ are the energy densities of radiation and non-relativistic fluid respectively, which satisfy the usual equations
\bea
\dot\rho_{m}=-3H\rho_{m}\,,\qquad \dot\rho_{r}=-4H\rho_{r}\,.
\eea
Finally, I  have the Klein-Gordon equation for $\psi$, which, in terms of $\phi$, becomes
\bea\label{KG}
\dot\phi+3H\phi+{6\eta \phi H\dot H\over \alpha+3\eta H^{2}}=0\,.
\eea
One can easily check that eq.\ \eqref{fr2} can be obtained as a linear combination of the derivative of eq.\ \eqref{fr1} and the remaining ones. The Klein-Gordon equation \eqref{KG} can be integrated explicitly to give the relation
\bea\label{phiH}
\phi={q\over a^{3}(\alpha+3\eta H^{2})}\,,
\eea
where $q$ is an arbitrary integration constant. 

Note that the modified Friedmann equation \eqref{fr1} takes the same form as the standard one if I  make the redefinition
\bea
\tilde\kappa=\kappa-\frac34 \eta\phi^{2}\,,
\eea
which, in terms of the Newton constant, corresponds to 
\bea
\tilde G={G\over 1-12\pi \eta \phi^{2}G}\,.\label{Geff}
\eea
In other words, the model \eqref{model} predicts a running Newton constant $\tilde G$ that, eventually, coincide with the ``bare'' value $G$ whenever $\phi$ vanishes (i.e. $\psi$ become constant). We stress that, as in most modified gravity theories, $\tilde G$ does not necessarily coincide with the value of the Newton constant that is measured, for instance, in Cavendish-like experiment or via Solar System tests. For the latter, in particular, one should analyse spherically symmetric solutions to the theory \eqref{model}. In some cases, the effects are null, as shown in the stealth neutron star solutions studied in \cite{HornNS}. In these models, when the scalar field is linear in time, the metric outside the distribution of matter coincide with the Schwarzschild metric. 

The above expression can be rewritten in the convenient form
\bea
{\tilde G\over G}={3\beta+\alpha \Omega_{\Lambda}\over 3\beta(1-\Omega_{\phi})+\alpha \Omega_{\Lambda}}\,,\label{GG}
\eea
where I  have defined the dimensionless parameter
\bea\label{beta}
\beta=\eta \Lambda\,,
\eea
and the relative energy densities associated to the scalar field ($\Omega_{\phi}$) and to the cosmological constant ($\Omega_{\Lambda}$), defined in eqs.\ \eqref{dens} below. It can also be shown that the dynamics depend on $\beta$ but not on $\Lambda$ and $\eta$ separately.  In the case $\alpha\neq 0$ (i.e. when one can normalise it to be $\alpha=1$), one sees that a very large $\beta$ implies that $\tilde G/G=(1-\Omega_{\phi})^{-1}$, which is the same as the case $\alpha=0$. On the other hand, when $\beta$ is small $\tilde G/G$ is basically constant for any $\alpha$. For $\alpha=0$, the Newton constant is still time dependent but independent of $\beta$. Finally, I  observe that, for $\alpha\neq 0$, the ratio $\tilde G/G$ tends to one in both the radiation dominated and dark energy dominates phases thus the total variation of $\tilde G$ is vanishing.

By using eq.\ \eqref{phiH}, one finds that eq.\ \eqref{fr1} can be written in the standard form
\bea\label{eqomega}
1=\Omega_{m}(t)+\Omega_{r}(t)+\Omega_{\Lambda}(t)+\Omega_{\phi}(t)\,,
\eea
where
\bea\label{dens}
\Omega_{m}={\rho_{m}\over 6\kappa H^{2}}\,,\quad \Omega_{r}={\rho_{r}\over 6\kappa H^{2}}\,,\quad \Omega_{\Lambda}={\Lambda\over 3H^{2}}\,,\quad \Omega_{\phi}={q^{2}(\alpha+9\eta H^{2})\over 12\kappa H^{2}a^{6}(\alpha+3\eta H^{2})}\,,
\eea
are the relative energy densities of matter, radiation, vacuum energy, and scalar field respectively.

The last component is the contribution to the overall energy density of the scalar field and it never vanishes if $q\neq 0$. In the next sections I  will show that this component can play the role of the pure dark matter fluid while $\Omega_{m}$ refers to baryonic matter only. The other possible scenarios are ruled out by the  comparison of the standard $\Lambda$CDM model with the numerical computation of eq.\ \eqref{eqomega}.

\section{The scalar field as cold dark matter}\label{sec3}

\begin{figure}[ht]
  \centering 
  \includegraphics[scale=0.9]{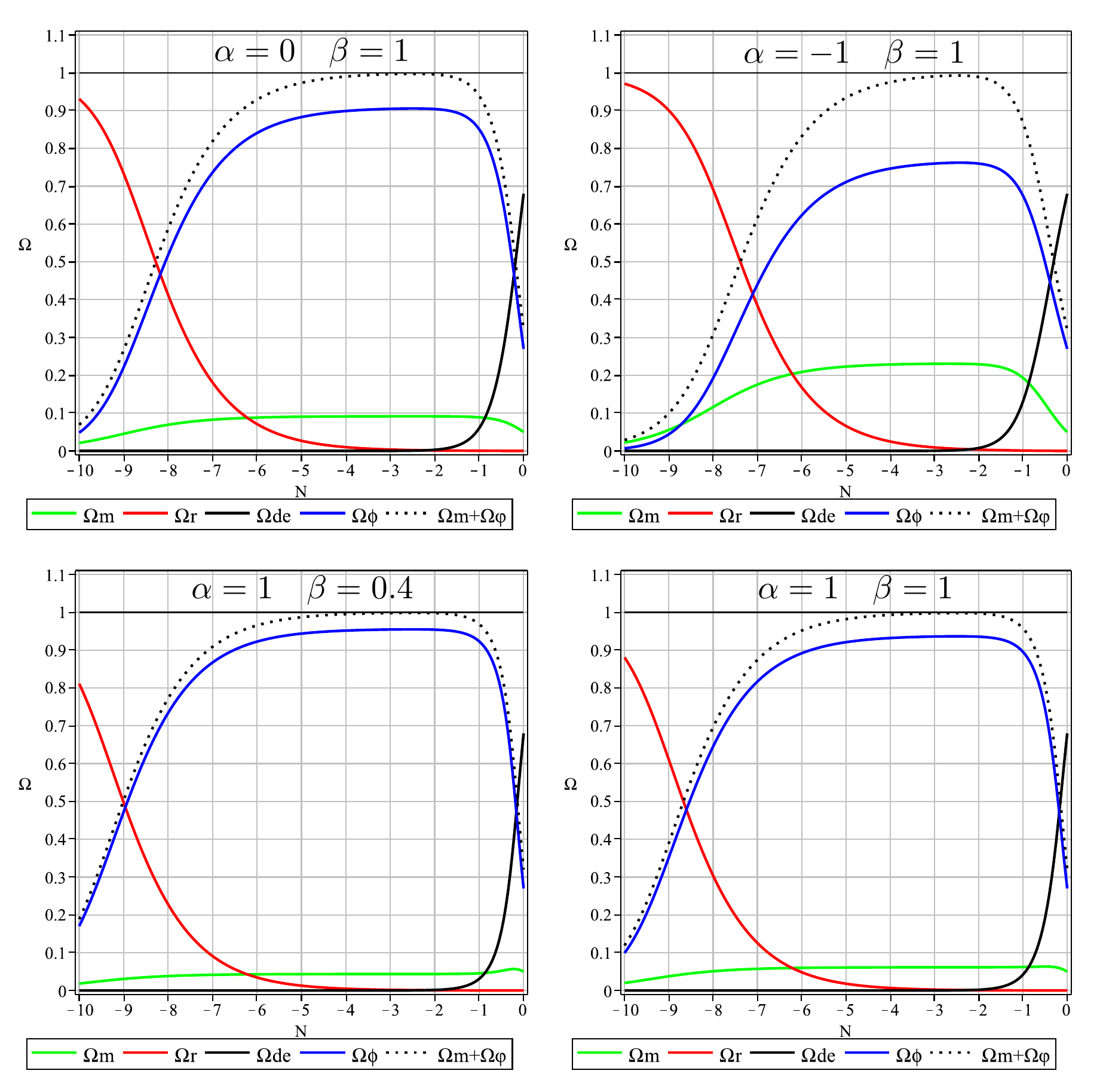} 
  \caption{Evolution of the relative energy densities, as explained in the text.}
  \label{allomega}
  \end{figure}

\noindent I  begin with the case where the scalar field acts as cold dark matter that adds up to ordinary baryonic matter. This is tantamount to setting the initial conditions today as
\bea\label{IC}
\Omega_{m}(0)=0.05\,,\quad \Omega_{r}(0)=10^{-4}\,\quad \Omega_{\Lambda}(0)=0.68\,,
\eea
and identifying $\Omega_{m}(0)$ as the current energy density of baryonic matter only. In fig.\ \ref{allomega} I  compare the evolution of the relative densities in four cases, two with $\alpha=1$ and the rest with $\alpha=0$ and $\alpha=-1$. In all cases I  consider $\beta>0$. In fact, I  found that the case $\alpha=-1$ and $\beta<0$ yields a negative $\phi^{2}$ so it will be discarder. On the other hand, the case $\alpha=1$ and $\beta<0$ is almost identically to the specular $\beta>0$. As pointed out before, the large $\beta$ case is very similar to the $\alpha=0$ case (see discussion after eq.\ \eqref{GG}). Therefore I  focus only on the cases presented in fig.\ \ref{allomega}. The energy densities are all computed with respect the efolding time $N=\ln a$, which is related to the redshift $z=a(0)/a(t)$ by the relation $N=-\ln (z+1)$. Red, blue, green, and black lines correspond to the energy density of radiation, dark matter, baryonic matter, and cosmological constant respectively. The dotted line is the sum of dark and baryonic matter. One notes that, in all plots the baryonic matter, is always subdominant with respect the dark matter component. The ratio between the two matter components is quite sensitive to the value of $\beta$ when this is of order one. The same holds for the matter/radiation equality time which moves according to the value of $\beta$.

\begin{figure}[ht]
  \centering 
  \includegraphics[scale=0.37]{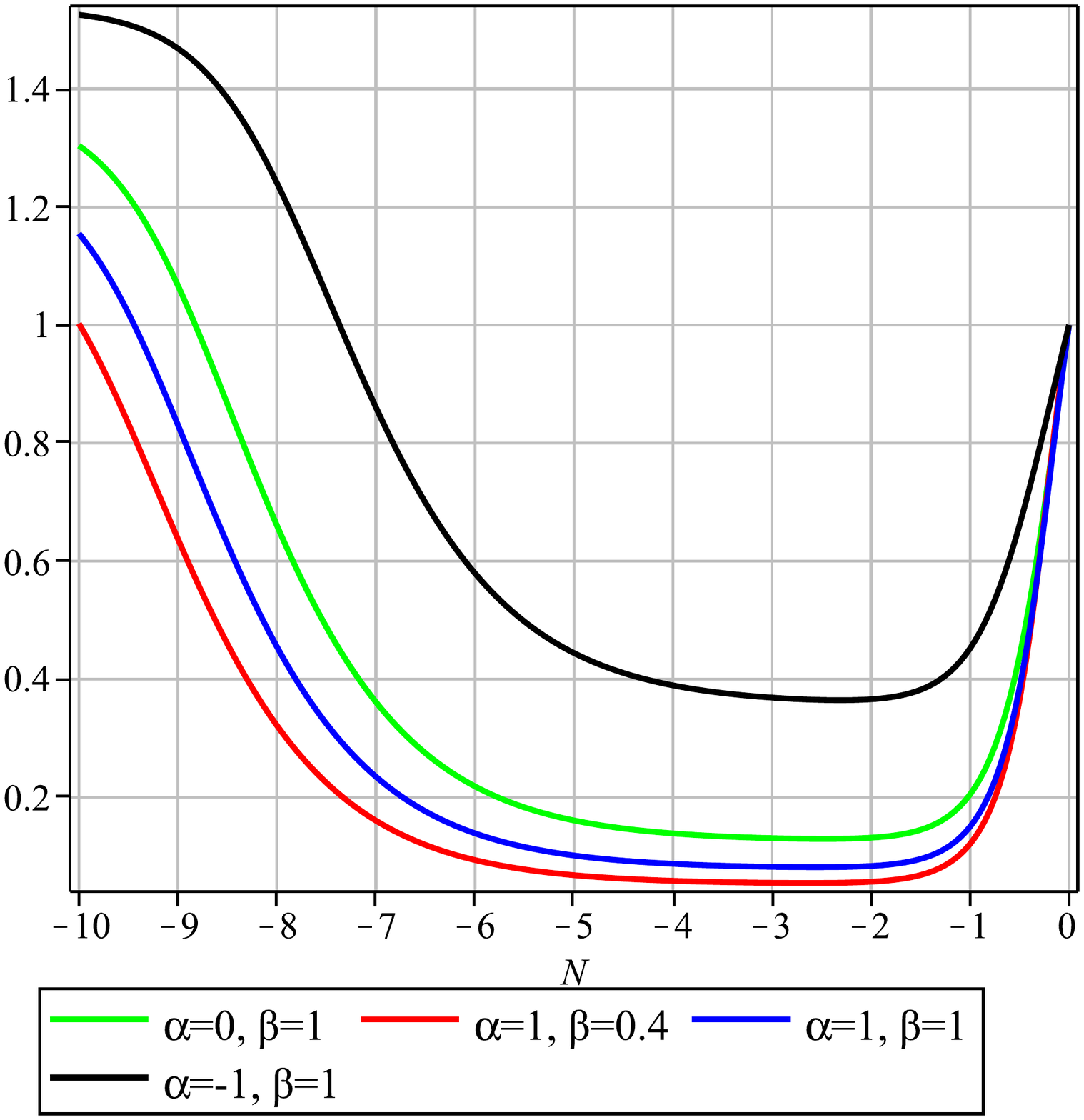} 
  \includegraphics[scale=0.37]{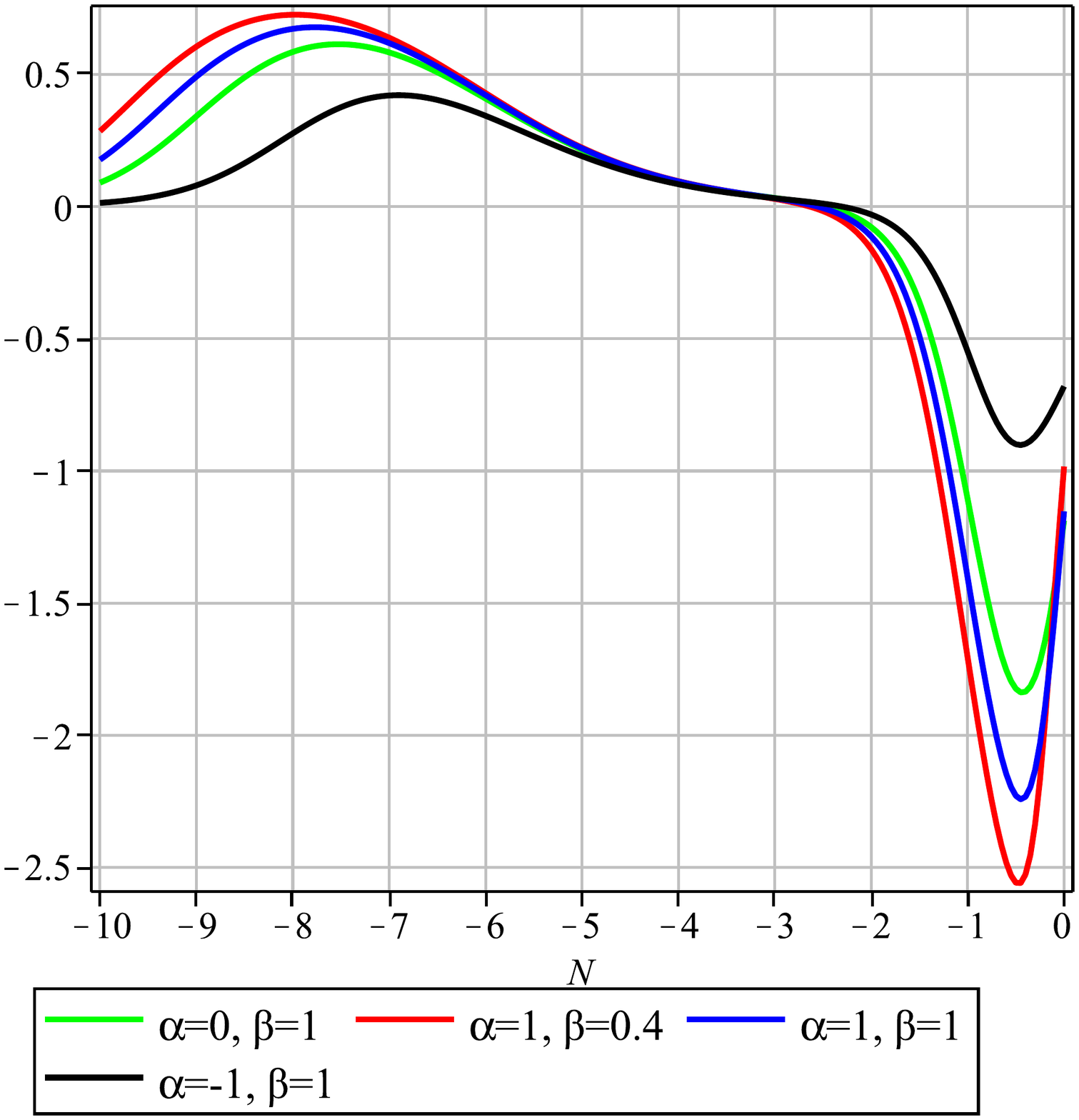} 
  \caption{Plots of $\tilde G/G$ (left) and  $\tilde G^{-1}d\tilde G/dN$ (right) as functions of $N$.}
  \label{GGplot}
  \end{figure}

It is interesting to study the time-evolution of the effective Newton constant, according to \eqref{GG}. In fig. \ref{GGplot}, I  show $\tilde G/G$ and $\tilde G^{-1}d\tilde G/dN$ as a function of $N$ for the same choices of $\alpha$ and $\beta$ as in fig.\ \ref{allomega}. One sees that the curves tend to converge at early and late times, as anticipated in the previous section. The time variation of $\tilde G$ on cosmological times should be compared with constraints coming from observations \cite{uzan}. For example, if we push our numerical calculation at $N\sim -19$, namely at the time of nucleosynthesis, we find that the time variation of $\tilde G$ is negligible and the observational constraints are satisfied.  We defer a full analysis to future work.

\begin{figure}[ht]
  \centering 
  \includegraphics[scale=0.45]{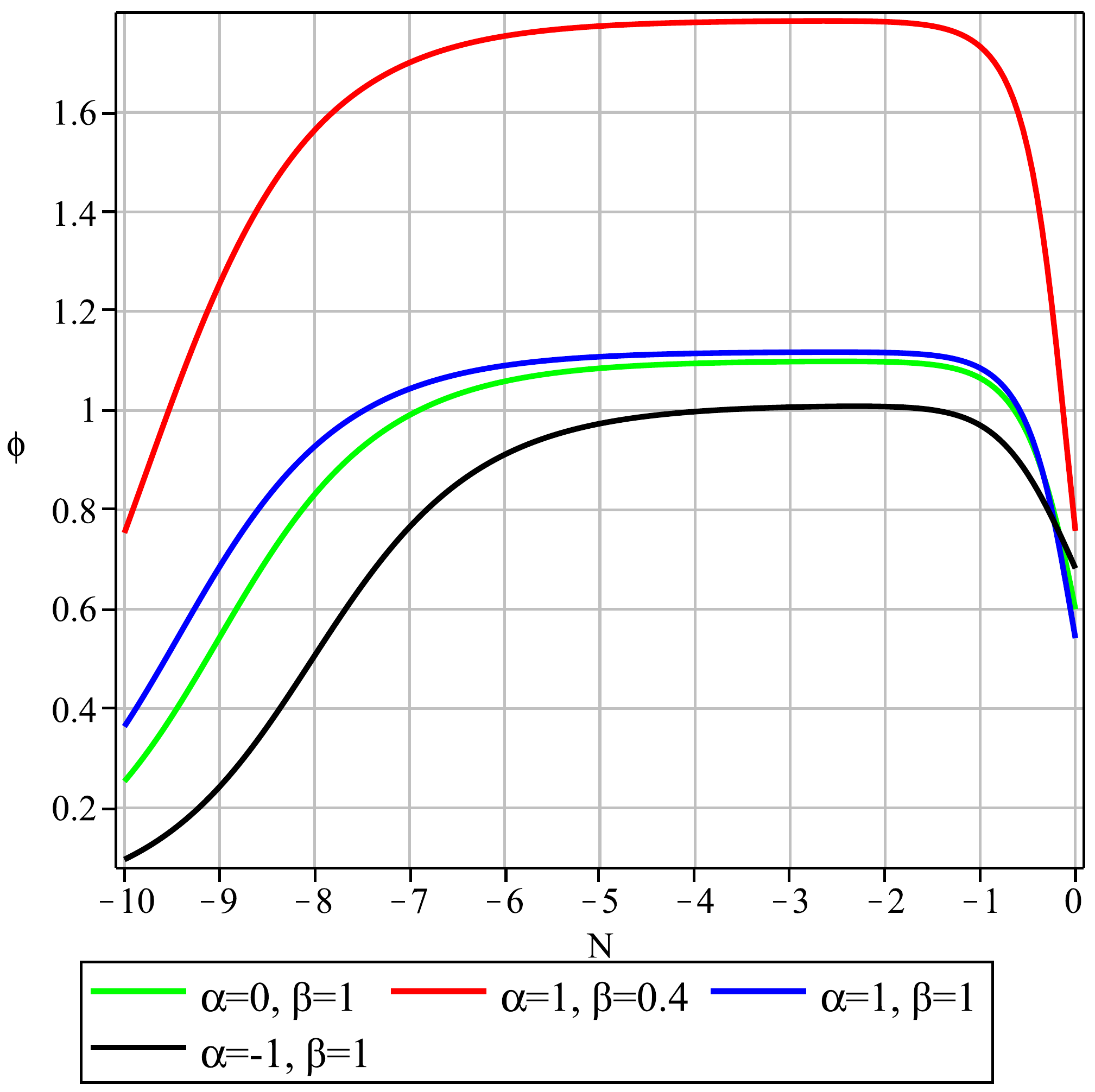} 
  \caption{Evolution of $\phi(N)$.}
  \label{phiplot}
  \end{figure}

To complete our analysis I  also study the equation of the scalar field $\phi=\dot \psi$. In some black hole solutions, the field $\phi$ can become imaginary in certain regions of spacetime (see e.g. \cite{HornBH} and \cite{greek}). In fact, this is not a big problem because, in those cases, the physical degree of freedom is $\phi^{2}$ and not $\phi$, as it is apparent from the equation of motions. In the cosmological model at hand instead, one has an implicit solution of $\phi$ in terms of the scale factor and the Hubble parameter. Thus one needs to check that it is real for all times. In fig.\ \ref{phiplot} I  plot the function $\phi(N)$ for the same parameter choices as in fig.\ \ref{allomega} and one sees that $\phi$ is real and positive at all times. In particular, one notes that during the radiation domination $\phi$ is almost constant while it become almost linear in $N$ in both early and recent Universe.

\begin{figure}[ht]
  \centering 
  \includegraphics[scale=0.8]{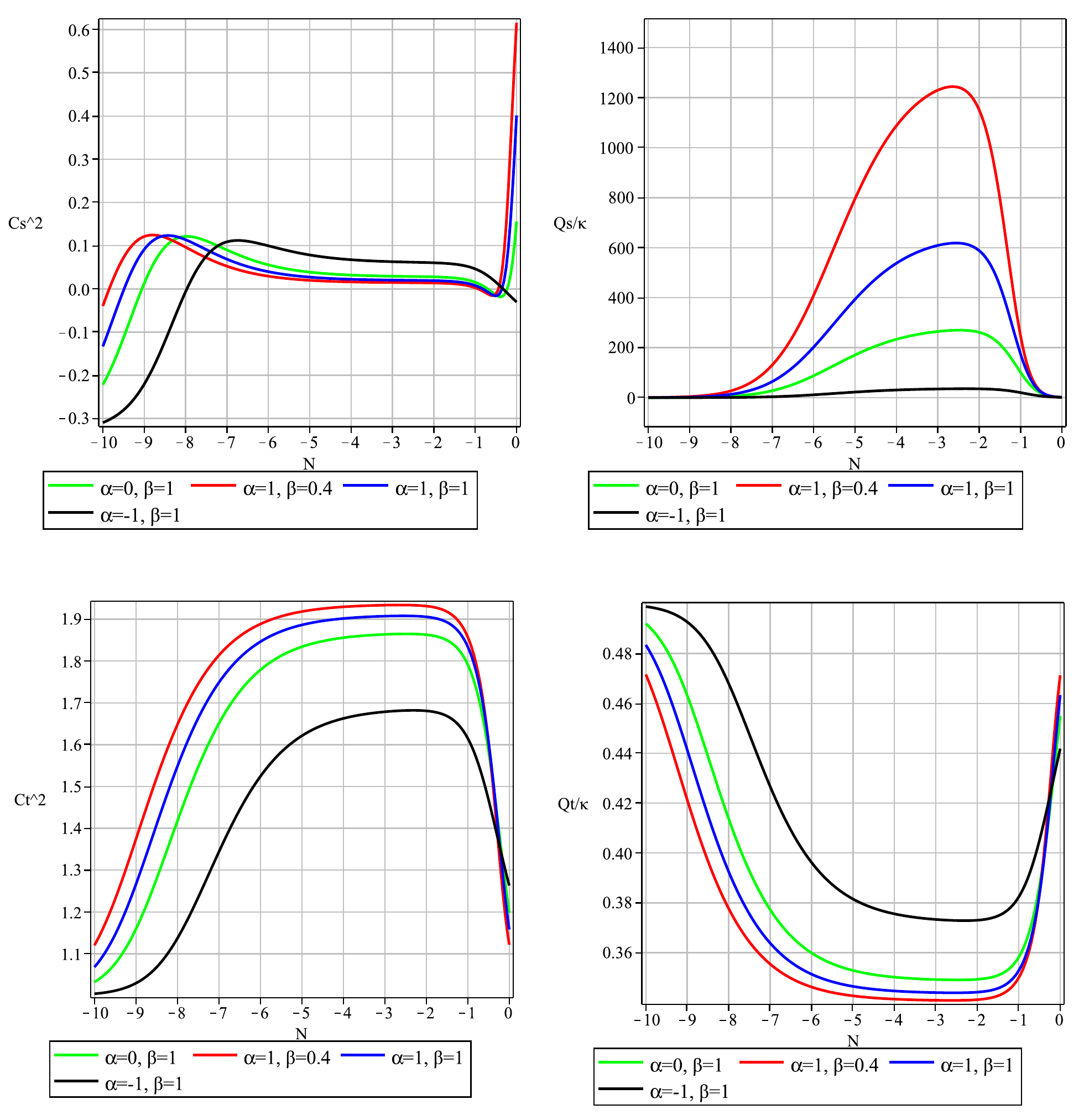} 
  \caption{Time evolution of $c_{S}^{2}$ (top left), $c_{T}^{2}$ (bottom left), $Q_{S}/\kappa$ (top right), and $Q_{T}/\kappa$ (bottom right).}
  \label{perturb}
  \end{figure}

We conclude this section by considering cosmological perturbations and, in particular, the presence of eventual instabilities. As shown in the appendix, in the context of Horndeski gravity there are general formulae that allows to verify the absence of ghost and Laplacian instabilities  \cite{tsudefel}. The conditions to be satisfied are that the squared speed of scalar and tensor perturbations, $c_{S}^{2}$ and $c_{T}^{2}$ respectively, together with the functions $Q_{S}$ and $Q_{T}$ defined in the appendix are positive (see eqs.\ \eqref{spert} and \eqref{tpert}). In fig.\ \ref{perturb} I  plot these functions for the same parameter choice as in fig.\ \ref{allomega}. One sees that they are all positive, except for the squared scalar speed of sound that appears to be slightly negative for a relatively short period around $N=-0.5$ ($z=0.6$). It would be interesting to understand whether this tiny violation has observable consequences that can be used to test (or rule out) model. In principle, in fact, this violation could have catastrophic effects. However, at the time the violation occurs large structures have already formed so it might not be relevant for the present model. 

\section{The scalar field is neither dark energy nor dark matter}\label{sec4}

\begin{figure}[ht]
  \centering 
  \includegraphics[scale=0.37]{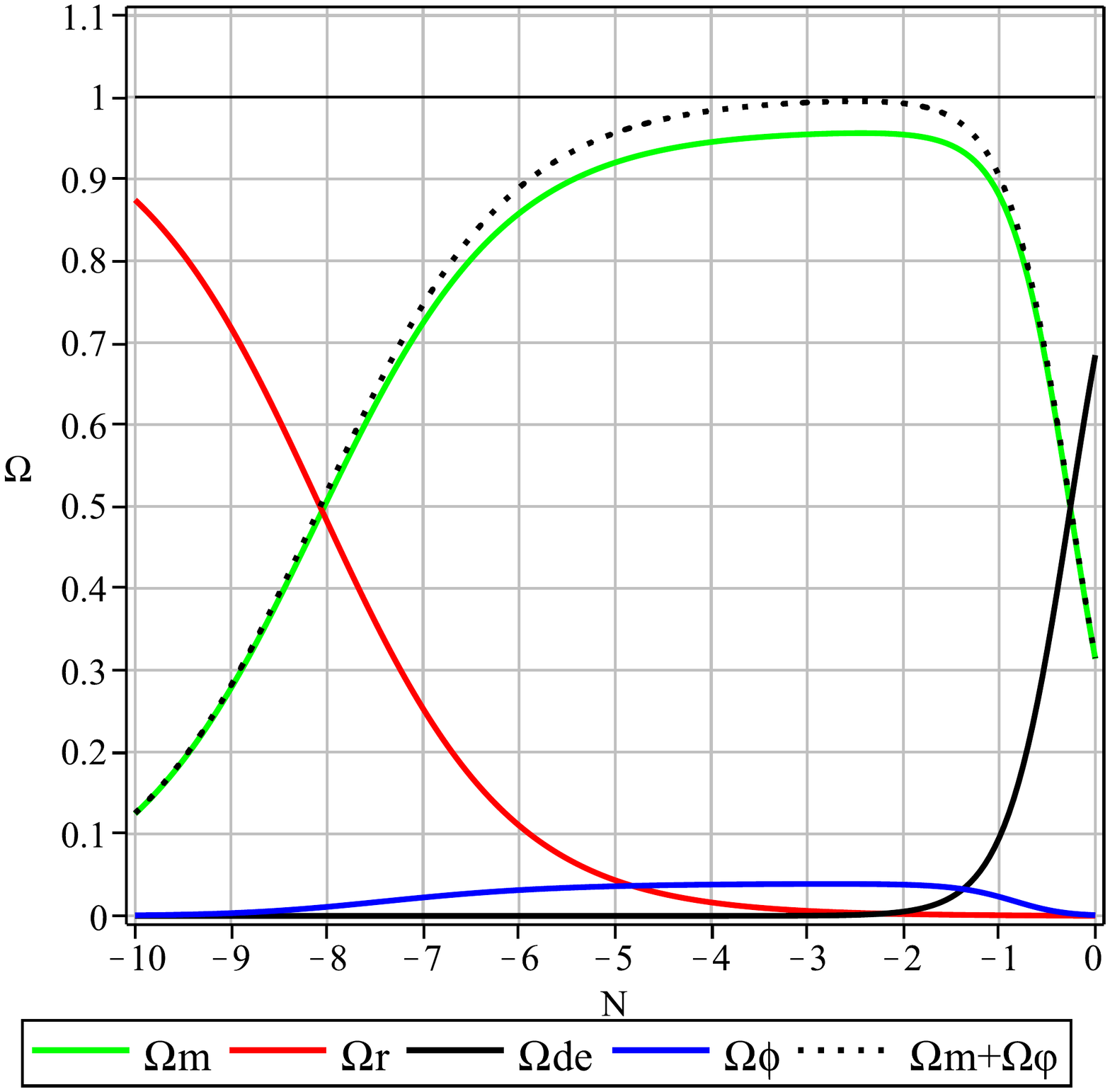} 
  \includegraphics[scale=0.37]{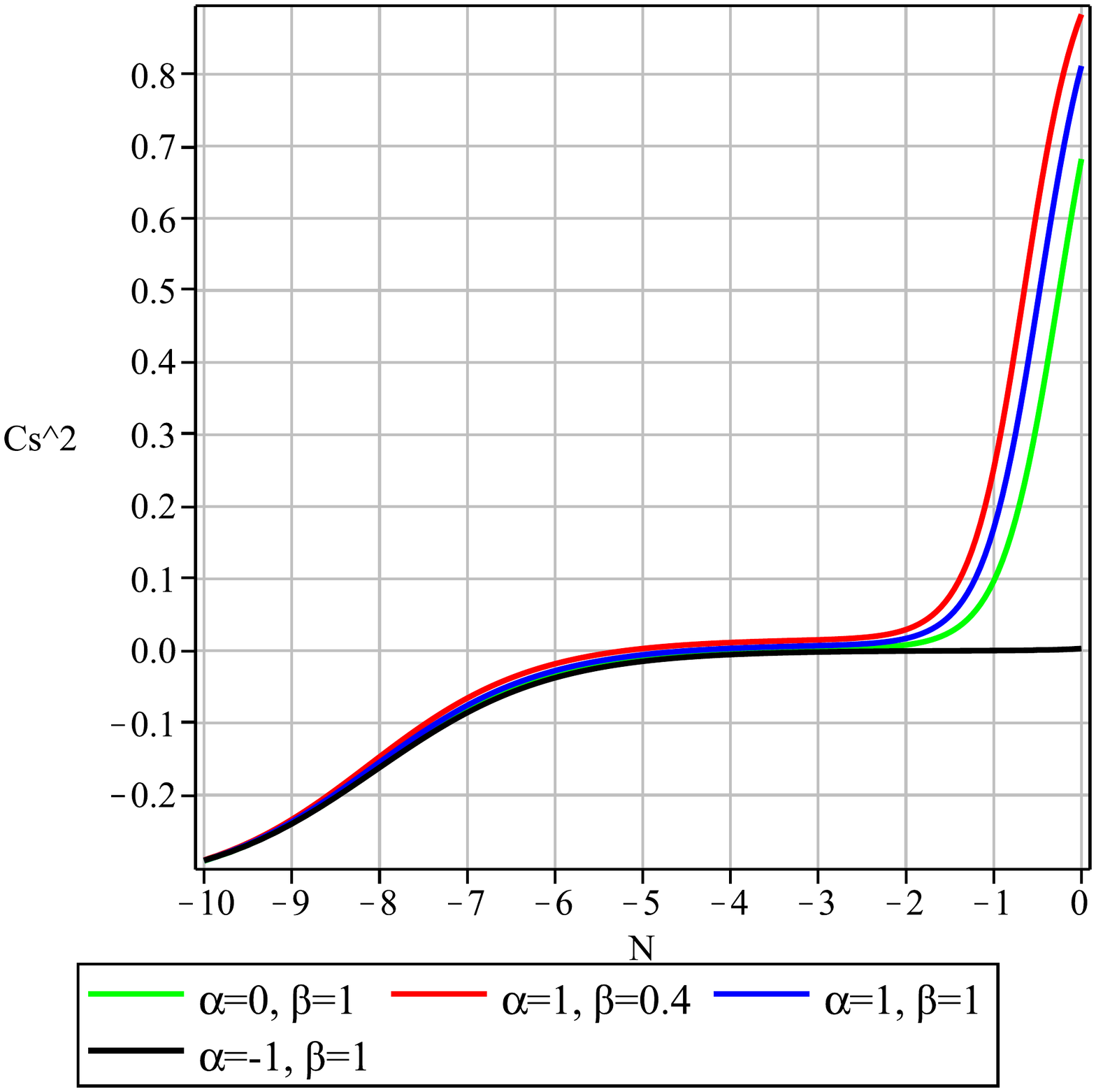} 
  \caption{On the left the plots of the relative energy densities for $\alpha=1$ and $\beta=0.4$. On the right the squared scalar speed of sound (right) in the case when the scalar field does not mimic dark matter.}
    \label{NoDM}
  \end{figure}

\noindent I  now consider the case when the scalar field is not dark matter nor dark energy. This implies that its energy density today has to be negligible, therefore I  choose the conditions
\bea
\Omega_{m}(0)=0.315\,,\quad \Omega_{r}(0)=10^{-4}\,\quad \Omega_{\Lambda}(0)=0.684\,.
\eea
For this case, I will show only one plot for the energy densities in the right panel of fig.\ \ref{NoDM}, where it is apparent how the contribution of the scalar field remains small through the cosmological ages. Most importantly, however, the model is affected by a large Laplacian instability at early time ($N<-5$) as it is clearly evident from  the left panel of fig.\ \ref{NoDM}. It follows that the model is unstable when its weight, in terms of energy density, is small. 

As a final note, I  observe that as $\beta$ increases, the contribution of the $\phi$ field in the evolution of the Universe becomes smaller, independently of the initial conditions. This is in line with what was found in a completely different context, namely black holes \cite{HornBH}. In that case it was explicitly proven that $\eta=\beta/\Lambda$ is in fact a non perturbative parameter and one could recover the exact Schwarzschild solution only for $\eta\rightarrow \infty$. 
  
\begin{figure}[ht]
  \centering 
  \includegraphics[scale=0.28]{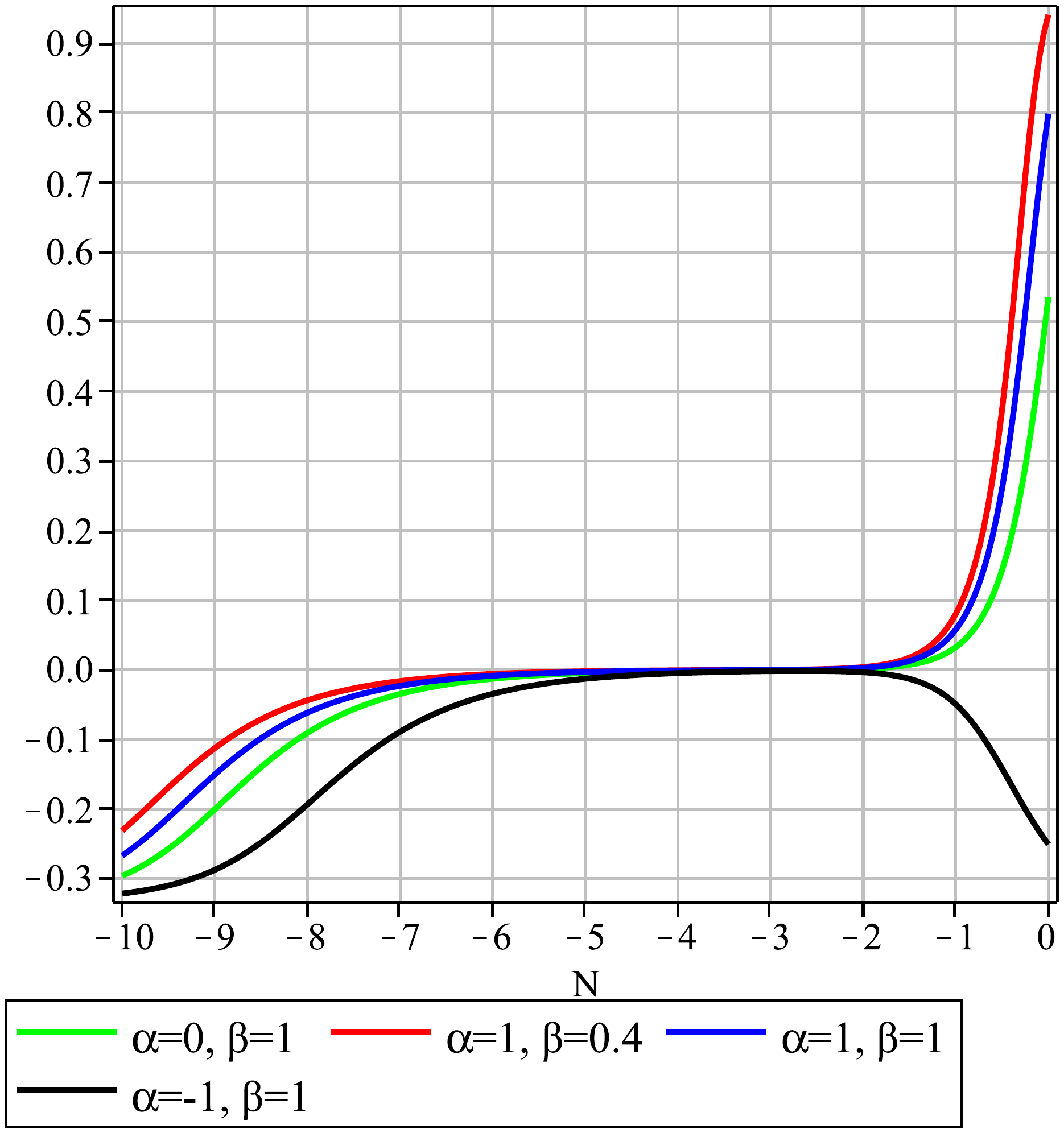} 
  \includegraphics[scale=0.28]{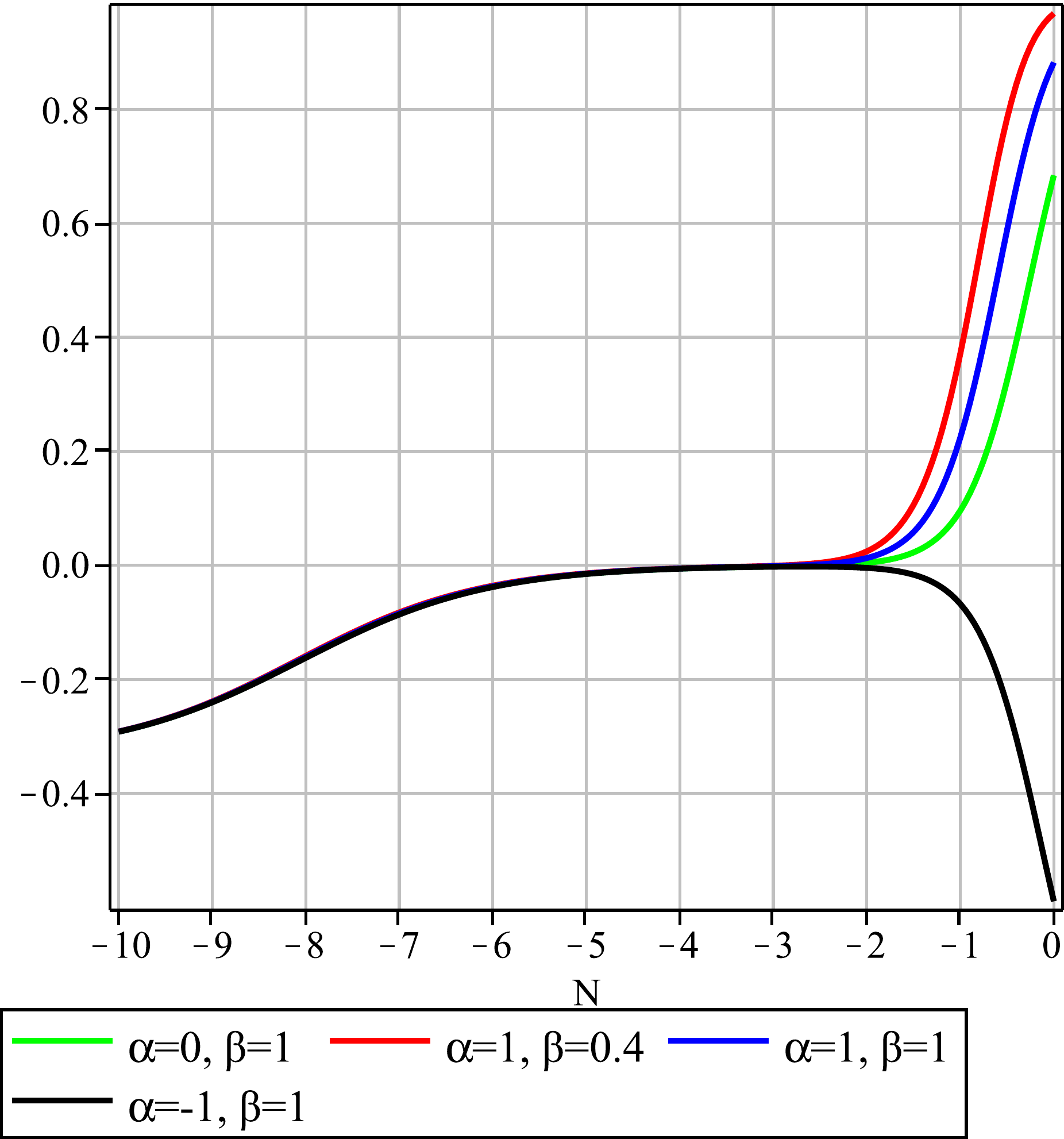} 
   \includegraphics[scale=0.28]{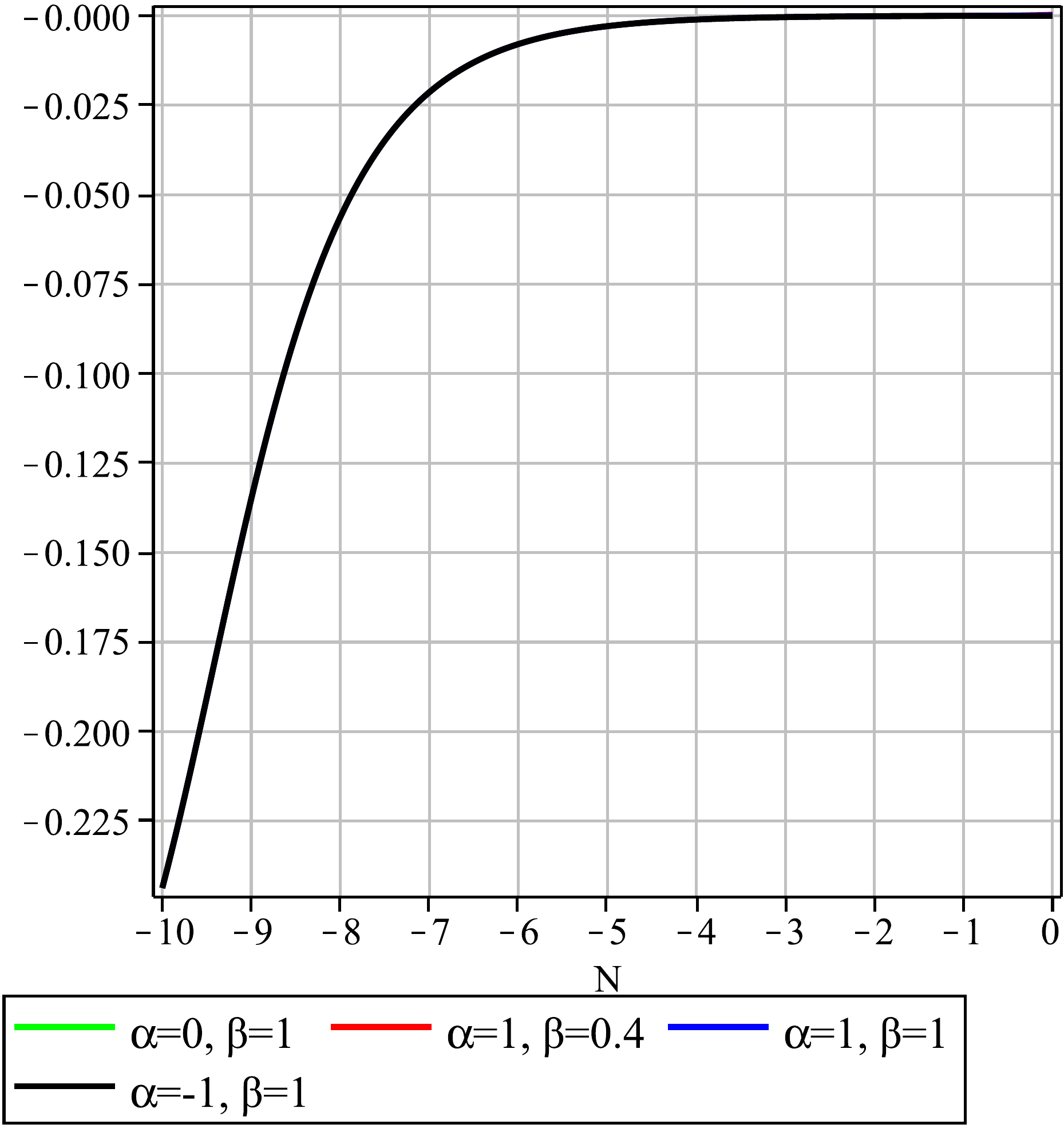} 
  \caption{Evolution of the effective equation of state relative to the scalar field \eqref{eos} in the three cases considered in the text.}
    \label{EoS}
  \end{figure}

To complete our analysis I  should also consider the possibility that the scalar field act as dark energy. However, this hypothesis can be ruled out by studying numerically the equation of state parameter associated to the scalar field that can be constructed from the energy momentum tensor associated to $\psi$. In terms of $\phi$  the equation of state can be written as
\bea\label{eos}
\omega_{\phi}={(\alpha-3\eta H^{2})\phi-2\eta(\phi \dot H+2\dot \phi H)\over \phi (\alpha+9\eta H^{2})}\,.
\eea
In fig.\ \ref{EoS} I  plot the equations of state in the three cases: on the left I  consider the case when the scalar field mimics dark matter, at the center when it is neither dark matter nor dark energy and, on the right, when it mimics dark energy (i.e when the initial conditions at $N=0$ are $\Omega_{m}(0)=0.315\,,\Omega_{r}(0)=10^{-4}\,,\Omega_{\Lambda}(0)=0$ and the term $\Lambda$ in the Lagrangian is considered as a ``bare'' parameter only). From the plot on the right one sees that the equation of state is never smaller than $-1/3$, which is a necessary condition to be considered as dark energy. I  also note that the equation of state is basically the same for the all the chosen values of $\alpha$ and $\beta$. In the other two cases,  when $\alpha=-1$, $\omega_{\phi}$ becomes negative. In particular, in the case when the scalar field is not dark matter, the equation of state parameter becomes smaller than $-1/3$ in the recent Universe (central panel, fig.\ \ref{EoS}).

In conclusion, I  find that the dark matter component at cosmological scales can be played by a massless scalar field coupled to the Einstein tensor and with vanishing potential, in order to preserve shift invariance. As a crucial test to assess whether this model can actually replaced a fluid made of unknown, non-interacting matter particles, one needs to verify another important phenomenological aspect of dark matter, namely the flatness of the rotation curves of spiral galaxies.

  \section{Rotation curves}\label{sec5}
  
  \noindent Having established that, at cosmological scales, the scalar field of our model can mimic dark matter, one should verify if it can also explain, at least to some degree, the anomalous flat rotation curve of spiral galaxies. 
  In the following, I  assume the the metric has the spherically symmetric form
  \bea\label{sphmetric}
  ds^{2}=-M(r)dt^{2}+{dr^{2}\over N(r)}+r^{2}d\Omega^{2}\,.
  \eea
  It is important to remember that the Klein-Gordon equation associated to the field $\psi$ takes the form $\nabla_{\mu} J^{\mu}=0$, where $J^{\mu}$ is the current associated to the shift symmetry defined in \eqref{current}. 
  
 In \cite{HornNS} it was shown that the subclass of spherically symmetric configurations named ``stealth'' and characterised by the scalar field of the form $\psi(r,t)=Qt+f(r)$, with $f(r)$ arbitrary  \cite{Babichev:2013cya},  is such that the metric outside a massive object matches exactly the Schwarzschild one when $\alpha=\Lambda=0$. Therefore, in this case I  do not expect that the rotation curve is different from the Newtonian one \footnote{However, the fact the \emph{ inside} the matter distribution the metric is different should modify the geodesics equation for point particles orbiting inside the disc.}.
 
 It is interesting to see what kind of cosmological solution the stealth scalar field implies. Assuming that the radiation content of the Universe is negligible and setting, therefore, $\phi=Q$ and $\rho_r=0$ in the equations \eqref{fr1}-\eqref{phiH} it is possible to show that, for $|N|\ll 1$, one has the condition
 \bea\label{qlin}
 Q^2={12\kappa H_0^2\Omega_{\phi,0}\over \alpha+9\eta H_0^2}\,,
 \eea
 where the subscript zero indicates quantities calculated at the present time. Therefore, for small redshift, a scalar field linear in time is compatible with a Universe dominated by baryonic matter and the scalar field itself, provided the above condition holds. Of course, this approximation breaks at large redshift but it is good enough to study the rotation curves in the present Universe since galactic disks have formed at $z\sim 1$ i.e at $N\sim -0.69 $.

In the following, I  will  consider $\psi$ linear in time, i.e. the stealth configuration of \cite{Babichev:2013cya}. Therefore, in general one has the parameters $(Q,\Lambda,\alpha,\eta)$. If $Q=\Lambda=0$ and $\alpha=1$ one recovers the black hole solution studied in \cite{HornBH}. For non-vanishing $\Lambda$  one finds the solutions studied in \cite{adolfoBH}. In both cases,  one takes advantage of the fact the the current has only one non-vanishing component and the Klein-Gordon equation reduces to $J_{r}=K$. In the case $K=0$, the equation $J_{r}=0$ yields
\bea\label{NM}
N={(\alpha r^{2}+\eta)M\over \eta(M+rM')}\,,
\eea
 where, from now on, a prime denotes the derivative with respect to $r$. The $tt$ and $rr$ component of the Einstein equation can then be solved analytically and one finds
 \bea\label{marctan}
 M=1-{2m\over 3}+C_{1}{r^{2}\over \ell^{2}}+C_{2}{\ell\over r}\arctan{\left(r\over \ell\right)}\,,
  \eea
  where $\ell=\sqrt{\eta/\alpha}$, and
  \bea
  C_{1}={(1-\Lambda \ell^{2})\over 3\ell^{2}(3+\Lambda \ell^{2})}\,,\quad  C_{2}={(1+\Lambda \ell^{2})^{2}\over (1-\Lambda \ell^{2})(3+\Lambda \ell^{2})}\,.
  \eea 
  In order for $M$ to be definite for all $r>0$ one needs  $\ell$ to be real, which implies that $\eta$ and $\alpha$ have the same sign. This solution is asymptotically locally isomorphic to the de Sitter ($\Lambda\ell^{2}\gg 1$) or anti de Sitter metric ($\Lambda\ell^{2}\ll 1$).  
  
 For a spherically symmetric potential one can approximate the rotation squared velocity of a star in the galactic disc as $V^{2}=rM'$.  Then, by expanding $V^{2}$ for large $z=r/ \ell$ one finds for the solution \eqref{marctan}
  \bea
  V^{2}\simeq 2C_{1}z^{2}+{1\over z}\left({2m\over \ell}-{\pi C_{2}\over 2}\right)+{\cal O}(C_{2}z^{-2})\,.
  \eea
  From this one clearly sees that, even if $C_{1}$ is negligible compared to $C_{2}$ (which happens whenever $\Lambda \ell^{2}\sim 1$), the squared velocity remains proportional to $r^{-1}$ as in Newtonian gravity. Therefore, this model is not able to describe the flattening of the galactic rotation curves.
  
The last possibility is that  $\alpha\neq 0$, $Q\neq 0$, and $\Lambda\neq 0$.  In such a case, one can show that the orbital velocity around a point mass matches the usual Keplerian profile at short and large distances while, at intermediate range, the slope increases. 

By allowing $Q\neq 0$, the $tr$ component of the Einstein equation vanishes if, and only if, eq.\ \eqref{NM} holds. It follows that the Klein-Gordon equation vanishes identically, and the combination of the $rr$ and $tt$ components of the Einstein equation yields a very complicated differential equation in $M$ only. However, by defining the function $W(r)$ such that
\bea\label{MW}
M(r)={\int (\alpha r^{2}+\eta)W(r)dr\over \eta r}\,,
\eea
the equation for $M$ can be written in the simpler form
\bea\label{eqW}
W'={4[(2\Lambda\eta+2)W-s]rW\eta\over 2(r^2+\eta)(-r^2+\eta r^2\Lambda-2\eta)W+3\eta s (r^2+\eta)} \,,
\eea
where I have set $\alpha=1$ and 
\bea\label{spar}
s={\kappa\over \eta Q^2}\,.
\eea
By differentiating the expression $V(r)=\sqrt{rM'}$, and with the help of the equations above, one obtains the equation
\bea\label{eqV}
2\eta r V V'+\eta V^{2}-2 r^{2}W-r( r^{2}+\eta)W'=0\,.
\eea
The equation for $W(r)$ can be solved implicitly, yielding  a cubic polynomial in $W$ but inspection of eq.\ \eqref{eqW} clearly shows that $W(r=0)= W_{0}$, for $W_{0}$ arbitrary. However, to match the Schwarzschild metric at small distances, one needs to set $W_{0}=1$, as it is apparent from the definition \eqref{MW}. On the other hand, for large $r$ one finds that $W\rightarrow 0$. Therefore,  $W$ maps $r\in[0,\infty]$  into the interval $[1,0]$. By solving \eqref{eqV} for $W=1$ one finds
\bea
V=\sqrt{{C\over r}+{2r^{2}\over 3\eta}}\,,
\eea
where $C$ is an integration constant. In the limit $r^{2}\ll\eta $ one recovers the usual profile $V\sim r^{-1/2}$. On the other hand, when, $W=0$ at large $r$, the solution is exactly
\bea
V=\sqrt{C\over r}\,,
\eea
as for small $r$. Therefore, it is proven that for both small and large distances the orbital velocity is Keplerian. If there is any flattening of the curve $V(r)$, it has to happen at some intermediate range. I assume that the scale $\sqrt{\eta}$ is much larger than the typical scale at which galactic rotation curves start to flatten out (about $10$ Kpc). Thus one can first solve eqs.\ \eqref{eqW} as a power series in  $r/\sqrt{\eta}\ll1$. One then substitute the result into eq.\ \eqref{eqV} and solve the equation for small $r/\sqrt{\eta}$. After some algebra one finds 
\bea
{V(r)\over c}=\left(r_{s}\over r\right)^{1/2}+{s+4\beta\over 3(3s-4)}\left(r\over r_{s}\right)^{1/2}{r^{2}\over \eta}+{\cal O}\left(r^{2}\over \eta\right),
\eea
where $c$ is the speed of light, $r_{s}$ is the Schwarzschild radius of the galaxy and $\beta$ is the parameter defined in eq.\ \eqref{beta}. The first term is the usual Keplerian velocity of a point particle orbiting a mass of Schwarzschild radius $r_{s}$. The second term is the first correction to the Keplerian velocity and grows with the distance. This shows that the velocity changes from decreasing to increasing. It is important to note that the slope of the curve determined by the second term is not universal since it depends on $r_{s}$ and this could explain the fact that different galaxies have slightly different rotation curves.

If one chooses $\eta\simeq \Lambda^{-1}$ (i.e. $\beta\sim 1$) as in the cosmological case, then the second term is subdominant with respect the first since
\bea
\left(r_{s}\over r_{\rm flat}\right)^{1/2}\simeq 3.5\times 10^{-5}\quad {\rm and} \quad\left(r_{\rm flat}\over r_{s}\right)^{1/2}{r_{\rm flat}^{2}\over \eta}\simeq7\times 10^{-8}
\eea
for a typical spiral galaxy like M31, for which $r_{s}\simeq 5\times 10^{11}$ m, and where $r_{\rm flat}$ is the typical distance at which the velocity curve starts to flatten out ($\geq 20$ Kpc), and $V(r_{\rm flat})/c\simeq 8\times 10^{-4}$. The factor ${(s+4\beta)/ (3(3s-4))}$ can be large when $s\rightarrow 4/3$. Only in this case then the correction to the rotation velocity can be large even when the scale $\sqrt{\eta}$ is comparable to the Hubble horizon.
On the other hand, if one assumes instead that the scale $\sqrt{\eta}$ is much smaller than the Hubble ratio, then correction can be larger and fit the data. It seems clear that, apart from the fine-tuned case $s\sim 4/3$, the scale $\eta$ cannot accommodate both cosmological and galactic dynamics: if one chooses $\sqrt{\eta}$ much smaller that the Hubble horizon then $\beta$ is vanishing and the cosmological equations reduce to a standard scalar-tensor theory with vanishing potential. 
Note that the combination of eqs.\  \eqref{spar} and \eqref{qlin} yields, for $\alpha=1$ 
\bea
\eta H_0^{2}=(12 s\Omega_{\phi,0}-9)^{-1}\,.
\eea
By assuming the initial conditions \eqref{IC} one has $\Omega_{\phi,0}\simeq 0.27$ and, hence, $s>2.78$ in order to have positive $\eta H^2_0$. Therefore, also the limiting case $s=4/3$ is ruled out, at least in the approximation $\phi=Q$. From the plots in fig.\ \ref{phiplot} it appears that $\phi$ is almost constant for $N<-2$ while it is linear in $N$ afterwards. This implies  that the gravitational potential, and hence the rotation curves, can change slightly in time right after the formation of the galactic disks. 

From these results one concludes that the possibility of mimicking dark matter both at cosmological and local scales exists if one assumes more scalar fields with derivative coupling. In any case, the precise analysis of the galactic rotation curves goes beyond the scope of this paper but I hope to have raised some interest in this model.
  
\section{Conclusions}\label{sec6}

\noindent In this paper I  have examined the possibility that dark matter can be simulated by a  modification of the Einstein-Hilbert action inspired by Horndeski gravity. I  have found that our model can mimic the standard $\Lambda CDM$  evolution at large scales for a range of values of the only parameter of the theory $\beta$ (for $\alpha=0,1$). I  also checked whether ghosts and Laplacian instabilities arise and I  found that the latter are can occur but for very short scales of time. I  finally checked the model at galactic scales and I  showed that it can accommodate the anomalous rotation curves but not with the same values of $\beta$ as in the cosmological models. However, if one assume that there are more scalar fields coupled to the Einstein tensor (with different coupling strengths), one can model both galactic dynamics and $\Lambda CDM$ at the same time.

\begin{acknowledgments}
\noindent I am grateful to S.\ Zerbini and L.\ Vanzo for fruitful discussions.
\end{acknowledgments}

\appendix*

\section{Ghosts and Laplacian instabilities}\label{app}

\noindent Here, I briefly recall the necessary conditions for stability in Horndeski gravity. I closely refer to the formalism laid out in ref.\ \cite{tsudefel}. The Horndeski Lagrangian, written in terms of the modern ``Galileon'' gravity, has the standard form 
\begin{equation}\label{Haction}
S=\sum_{i=2}^{5}\int d^{4}x\sqrt{-g}{\cal L}_i\ ,
\end{equation}
where
\bea
{\cal L}_2&=&G_2\ ,\\\non
{\cal L}_3&=&-G_{3}\square\phi\ ,\\\non
{\cal L}_4&=&G_{4}R+G_{4{X}}\left[(\square\phi)^2-(\nabla_{\mu}\nabla_{\nu}\phi)^2\right]\ ,\\\non
{\cal L}_5&=&G_{5}G_{\mu\nu}\nabla^{\mu}\nabla^{\nu}\phi-\frac{G_{5{X}}}{6}\left[(\square\phi)^3+2(\nabla_{\mu}\nabla_{\nu}\phi)^3\
-3(\nabla_{\mu}\nabla_{\nu}\phi)^2\square\phi\right]\ .
\eea
Here, $G_i$ are arbitrary functions of the scalar field and of its canonical kinetic term $X\equiv -\nabla_{\mu}\phi\nabla^{\mu}\phi\,$, while $G_{iX}$ denote their derivatives with respect to $X$. Our model does not belong directly to this classification. However, it can be easily shown that, by choosing $G_{2}=\alpha X$, $G_{3}=0$, $G_{4}=\kappa$, $G_{5}=-{\eta\psi/2}$ and by integrating by parts one obtains eq.\ \eqref{model}. By perturbing the action up to second order, one finds two conditions for the absence of ghost and Laplacian instabilities for scalar perturbations, namely
\bea\label{spert}
Q_{S}&=&{w_{1}(4w_{1}w_{3}+9w_{2}^{2})\over 3w_{2}^{2}}>0\,,\\
c_{S}^{2}&=&{ 3 ( 2 w_{1}^{2} w_{2} H - w^{2}_{2} w_{4} + 4 w_{1} w_{2} \dot{w_{1}}  - 2 w_{1}^{2}\dot{w2} )  - 6w_{1}^{2} [ ( 1 + \omega_{a} ) \rho_{a}  + ( 1 + w_{b} ) \rho_{b} ]   \over  w_{1}(4w_{1}w_{3}+9w_{2}^{2}) }\geq 0\,,\non
\eea
where $\omega_{a,b}$ and $\rho_{a,b}$ are the equations of state and the energy densities of two generic perfect fluids. If one of these is a cosmological constant, it does not contribute to the speed of sound. In our case,I considered the two fluids to be ordinary baryonic matter and radiation andI neglected the cosmological constant altogether.
For tensor perturbations, one has instead
\bea\label{tpert}
Q_{T}={w_{1}\over 4}>0\,,\qquad c_{T}^{2}={w_{4}\over w_{1}}\geq 0\,.
\eea
The functions $c_{S}$ and $c_{T}$ are the speed of sound of the scalar and tensor perturbations. The coefficients $w_{1..4}$ are obtained from the perturbed Lagrangian and their general expression can be found in \cite{tsudefel}. In our case, these expressions reduce to 
\bea
w_{1}&=&2\kappa-{\eta\over 2}\phi^{2}\,,\\
w_{2}&=&4\kappa H-3\eta H\phi^{2}\,,\\
w_{1}&=&{3\alpha\over 2}\phi^{2}-18\kappa H+27\eta H^{2}\phi^{2}\,,\\
w_{1}&=&2\kappa +{\eta\over 2}\phi^{2}\,.
\eea



%

\end{document}